\documentclass[a4paper]{article}

\usepackage{INTERSPEECH2020}

\usepackage{float}
\usepackage{multirow}
\usepackage{diagbox}
\usepackage{cite}

\title{Acoustic Echo Cancellation by Combining Adaptive Digital Filter and Recurrent Neural Network}
\name{Lu Ma, Hua Huang, Pei Zhao, Tengrong Su}
\address{Haier Smart Home Co., Ltd.}
\email{malu@haier.com, iamroad@163.com}

\begin{document}

\maketitle
\begin{abstract}
  Acoustic Echo Cancellation (AEC) plays a key role in voice interaction. Due to the explicit mathematical principle and intelligent nature to accommodate conditions, adaptive filters with different types of implementations are always used for AEC, giving considerable performance. However, there would be some kinds of residual echo in the results, including linear residue introduced by mismatching between estimation and the reality and non-linear residue mostly caused by non-linear components on the audio devices. The linear residue can be reduced with elaborate structure and methods, leaving the non-linear residue intractable for suppression. Though, some non-linear processing methods have already be raised, they are complicated and inefficient for suppression, and would bring damage to the speech audio. In this paper, a fusion scheme by combining adaptive filter and neural network is proposed for AEC. The echo could be reduced in a large scale by adaptive filtering, resulting in little residual echo. Though it is much smaller than speech audio, it could also be perceived by human ear and would make communication annoy. The neural network is elaborately designed and trained for suppressing such residual echo. Experiments compared with prevailing methods are conducted, validating the effectiveness and superiority of the proposed combination scheme.
\end{abstract}

\section{Introduction}
Acoustic echo will arise if the sound is listened by the speaker itself \cite{echo}. This phenomenon is very commonplace no matter in communications, entertainments or man-machine interaction, and somewhere else. It may be useful in some scenarios, such as entertainments. But, in most cases, especially for voice interactions and communications, it is interfering and should be cancelled from the significant speech audio \cite{aec}.

Since there is a reference signal representing the source of echo, adaptive filters are always employed for acoustic echo cancellation (AEC). There are many adaptive algorithms available, such as least mean square (LMS) \cite{LMS}, normalized LMS (NLMS) \cite{NLMS}, block LMS (BLMS) \cite{BLMS}, and etc. Each has its own merits and special applications. For obtaining considerable performance, filter lengths of several hundreds and sometimes thousands are required. Due to the significant reduction in computational load by using the fast Fourier transform (FFT) for implementing the block BLMS algorithm efficiently, the frequency domain block adaptive filter (FDBAF) based on the LMS algorithm is considered to be most suitable \cite{{PBFDAF}}. Moreover, for accommodating long block delay and large quantization error in FFT, a more flexible frequency domain adaptive filter structure, called the multidelay block frequency domain (MDF) adaptive filter was proposed \cite{MDF}. Further, for obtaining robust echo cancellation, methods for adjusting the learning rate to vary according to conditions such as double-talk and echo path change were also raised \cite{rate}. In brief, there are plenty of algorithms by using adaptive filter for AEC, giving considerable performance.

Unfortunately, there would be some residual echo after adaptive filtering. Though it is much smaller than speech audio in terms of amplitude in most cases, it could also be perceived by human ear and would make communication annoy. These residual echo includes linear residue introduced by mismatching between estimation and the reality and non-linear residue which are mostly caused by non-linear components on the audio devices. The linear residue can be reduced with elaborate structure and methods, such as \cite{rate} \cite{vss1, vss2, vss3, vss4}, leaving the non-linear residue intractable for suppression. Though, some non-linear processing (NLP) methods have already be raised, the algorithm processing are complicated and could be inefficient for suppression \cite{NLP3, NLP4}. Moreover, these NLP methods would bring damage to the speech audios \cite{NLP5}. In addition, some other methods such as non-linear filtering \cite{kalman} and modeling estimation \cite{modeling} are also used for non-linear echo cancellation.

By comparing the spectrum of residual echo with that of the speech audio, this residue can be considered as a type of noise. In addition, the far-end reference signal could also provide some relations for residue suppression. Inspired by this, a combination scheme by concatenating adaptive filter and neural network is proposed in this paper. The echo interfered speech audio is first processed by MDF filter with adaptive learning rate for cancelling primary echo signal. Thereafter, a neural network with perspicuous structure is elaborately designed and trained for residual echo suppression. This method is compared with other prevailing methods in terms of echo return loss enhancement (ERLE), logarithmic spectral distance (LSD), response time (RT), model size.

\section{Algorithm Structure}
\subsection{Combination Scheme}
\begin{figure}
\centering
\includegraphics[scale=0.63]{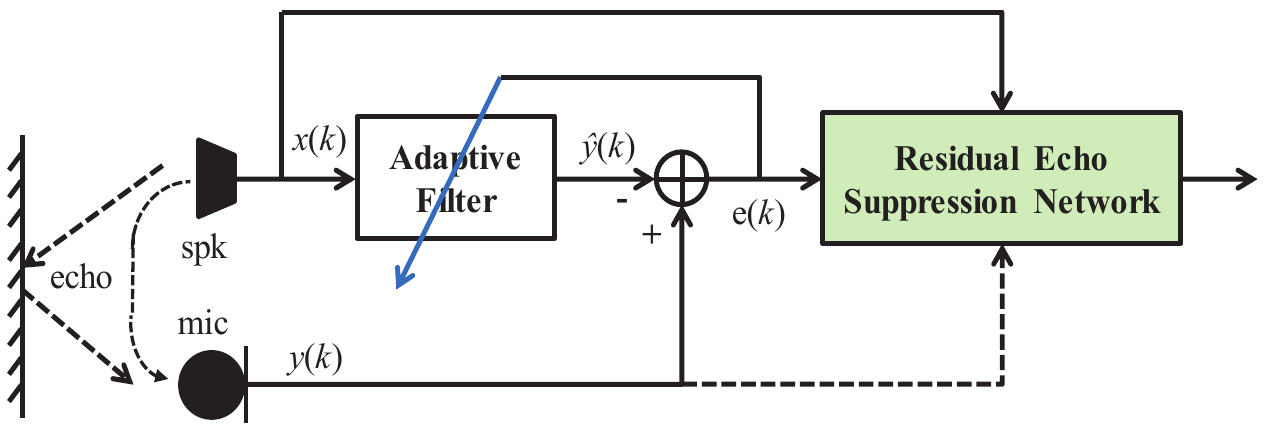}
\caption{Structure of combination scheme}
\label{fig:combine}
\end{figure}

\begin{figure}
\centering
\includegraphics[scale=1]{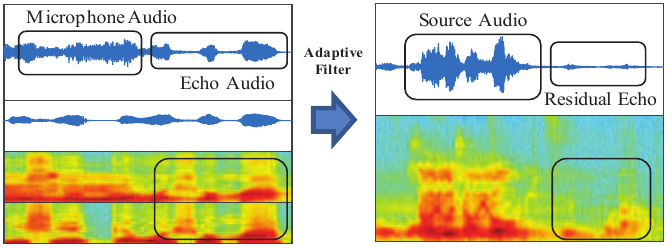}
\caption{Residual echo after adaptive filtering}
\label{fig:residue}
\end{figure}

The integration scheme by combining adaptive filter and the neural network is depicted in Fig. \ref{fig:combine}. Adaptive filter is used for cancelling the linear echo introduced by the multi-path or the room impulse response (RIR) \cite{RIR}. It has been proved to give considerable performance with low complexity. The weighting coefficients of the finite impulse filter (FIR) can be adjusted in time for estimating the RIR, then getting the estimated transcript of the echo signal. However, due to the non-linear components equipped on the devices, such as the loudspeaker with poor linearity, non-linear echo would be introduced. It cannot be cancelled by the adaptive filtering with FIR structure, resulting in residual echo. As is depicted in Fig. \ref{fig:residue}, the residual echo after adaptive filtering would be decreased to a little scale compared with speech audio in terms of amplitude. It could be considered as a special type of noise. Meanwhile, this noise could have some relations with the far-end reference signal. Therefore, based on these observations, a neural network will be designed and specialized trained for suppressing such residual echo as illustrated in Fig. \ref{fig:combine}.

\subsection{Adaptive Filter}
Due to the numerous merits involved in the multidelay block frequency domain adaptive filter, such as less memory storage, small FFT size, and allows different configurations to be chosen depending on the hardware used \cite{MDF}, it is employed in the combination scheme for linear echo cancellation. Moreover, as is used in the open source of Speex \cite{speex1, speex2}, the learning rate in the adaptive filter is controlled varying according to conditions such as double-talk and echo path change. In this case, the linear echo can be cancelled greatly and adaptively.

The complex NLMS filter of length $N$ is defined as
\begin{equation}
e\left( n \right) = d\left( n \right) - \hat y\left( n \right) = d\left( n \right) - \sum\limits_{k = 0}^{N - 1} {{w_k}\left( n \right)x\left( {n - k} \right)}
  \label{eq1}
\end{equation}
with adaptation step as
\begin{equation}
{\hat w_k}\left( {n + 1} \right) = {\hat w_k}\left( n \right) + \mu  \cdot \frac{{e\left( n \right)}}{{\sum\nolimits_{i = 0}^{N - 1} {\left| {x{{\left( {n - i} \right)}^2}} \right|} }} \cdot {x^ * }\left( {n - k} \right)
  \label{eq2}
\end{equation}
Where $x\left( n \right)$ is the far-end signal, $d\left( n \right)$ is the received microphone signal, $\hat y\left( n \right)$ is the estimated echo by adaptive filter,   $e\left( n \right)$ is the corresponding estimated error, ${w_k}\left( n \right)$ are the filter weights at time $n$, and ${\hat w_k}\left( n \right)$ is the estimated one, $\mu$ is the learning rate.

For obtaining a fast response in the case of double-talk in order to prevent the filter from diverging when double-talk starts, the learning rate is updated by \cite{rate}
\begin{equation}
{\hat \mu _{opt}}\left( {k,l} \right) = \min \left( {\hat \eta \left( l \right)\frac{{{{\left| {\hat Y\left( {k,l} \right)} \right|}^2}}}{{{{\left| {E\left( {k,l} \right)} \right|}^2}}},{\mu _{\max }}} \right)
  \label{eq3}
\end{equation}
where $\hat Y\left( {k,l} \right)$ and $E\left( {k,l} \right)$ are the frequency domain counterparts of the $\hat y\left( n \right)$ and $e\left( n \right)$, and $k$ is the frequency index and $l$ is the frame index, $\hat \eta \left( l \right)$ is the estimate leakage coefficient that represents the misadjustment of the filter. It is equal to the linear regression coefficient between the estimated echo power ${P_Y}\left( {k,l} \right)$ and output power ${P_E}\left( {k,l} \right)$:
\begin{equation}
\hat \eta \left( l \right) = \frac{{\sum\nolimits_k {{R_{EY}}\left( {k,l} \right)} }}{{\sum\nolimits_k {{R_{YY}}\left( {k,l} \right)} }}
  \label{eq4}
\end{equation}
where the correlations ${R_{EY}}\left( {k,l} \right)$ and ${R_{YY}}\left( {k,l} \right)$ are averaged recursively as:
\begin{equation}
\begin{array}{l}
{R_{EY}}\left( {k,l} \right) = \left( {1 - \beta \left( l \right)} \right){R_{EY}}\left( {k,l} \right) + \beta \left( l \right){P_Y}\left( k \right){P_E}\left( k \right)\\
{R_{YY}}\left( {k,l} \right) = \left( {1 - \beta \left( l \right)} \right){R_{YY}}\left( {k,l} \right) + \beta \left( l \right){\left( {{P_Y}\left( k \right)} \right)^2}\\
\beta \left( l \right) = {\beta _0}\min \left( {\frac{{\hat \sigma _{\hat Y}^2\left( l \right)}}{{\hat \sigma _e^2\left( l \right)}},1} \right)
\end{array}
  \label{eq5}
\end{equation}
where ${\beta _0}$ is the base learning rate for the leakage estimate and $\hat \sigma _{\hat Y}^2\left( l \right)$ and $\hat \sigma _e^2\left( l \right)$ are the total power of the estimated echo and the output signal. The variable averaging parameter $\beta \left( l \right)$ prevents the estimate from being adapted when no echo is present.

However, due to non-linear component involved in the device, non-linear residual echo will be arise at the output of adaptive filter. Moreover, some linear residual echo could be introduced if the estimated RIR and the actual one are mismatched. These would all result in considerable residual echo as depicted in Fig. 2. This residual echo would be more severe with the increasing nonlinearity introduced to the device and the increasing estimated error of the RIR.

\subsection{Neural Network}
\subsubsection{Network Structure}
\begin{figure}
\centering
\includegraphics[scale=0.5]{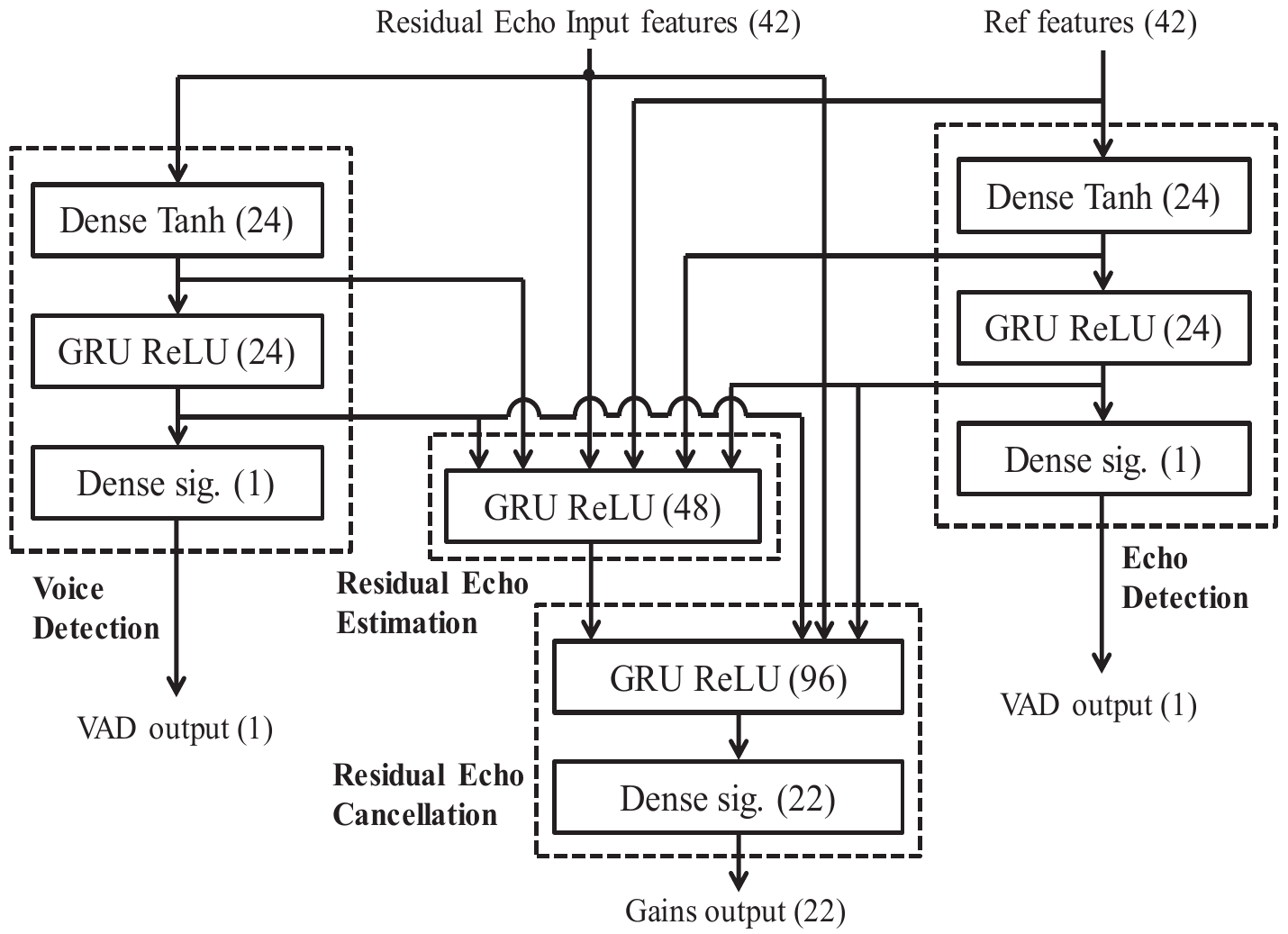}
\caption{Structure of the residual echo suppression network}
\label{fig:structure}
\end{figure}

Inspired by \cite{RNN}, the structure of residual echo suppression network based on Recurrent neural network (RNN) is elaborately designed and depicted in Fig. \ref{fig:structure}. Here, each module of RNN is realized by Gated Recurrent Unit (GRU) for data memory and network calculation. This type of structure mainly refers to the functional architecture of conventional echo cancellation, including three functional modules, i.e., double-talk detection, echo estimation and echo cancellation. Double-talk detection detects the signals of the far-end and near-end in real time, and only when a signal at the far-end is detected, would echo suppression be carried out. At this moment, residual echo is estimated from the signal after adaptive filtering. Echo cancellation, estimating the gain of the subband, rapidly changes the level of each frequency band in order to attenuate the echo but allow the signal to pass through. The reason for utilizing subband gain for computation is that it makes the model very simple, requiring very few band calculations. In addition, there are no so-called musical noise artifacts.

\begin{enumerate}
\item \textbf{Feature extraction.} In order to reduce the number of neurons thus reducing the model size, samples or spectrum are not directly used. Instead, the frequency band with bark scale is employed, matching the human perception. In this case, a total of 22 frequency subbands are used, namely, bark-frequency cepstral coefficients (BFCC). In addition, the first-order and second-order differences of the first six BFCC features, the discrete cosine transform (DCT) of the first six pitch correlation coefficients, and the dynamic features, i.e., pitch period and spectral non-stationarity metric were extracted \cite{RNN}. These all result in 42 features in total, acting as the input data of residual echo suppression neural network.

\item \textbf{Double-talk detection (DTD).} Only speech signal together with residual echo would be reserved after adaptive filtering. Since the amplitude of residual echo after adaptive filtering is small, the voice activity of speech can be easily detected. Meanwhile, the voice activity of the reference signal from far-end can also be easily detected due to its purity. In this case, two voice activity detections (VADs) for each channel can be implemented independently, reducing the difficulty of DTD.

\item \textbf{Residual echo estimation.} As a realization of recurrent neural network (RNN), gated recurrent unit (GRU) module is used for estimating residual echo with input features of reference signal, output signal of adaptive filtering and the DTD results. Due to the memory function of RNN model, residual echo can be better estimated compared with other models.

\item \textbf{Residual echo suppression.} A GRU module concatenated by a dense layer is used for echo suppression by calculating subband gains. It will approach to zero if the VAD of the near-end, i.e., the output of adaptive filtering, is zero, and will approach to one, if the VAD of the reference signal from the far-end is zero. Otherwise, a decimals representing the ratio between the speech and that superimposed by residual echo is estimated.
\end{enumerate}

\begin{figure}
\centering
\includegraphics[scale=1.5]{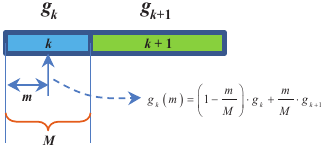}
\caption{Schematic of linear interpolation}
\label{fig:interpolation}
\end{figure}

Since only band gains are calculated by network, it can not be directly applied to each frequency. Therefore, linear interpolation between bands for obtaining frequency gain is required which is illustrated in Fig. \ref{fig:interpolation} and can be formulated as
\begin{equation}
{g_k}\left( m \right) = \left( {1 - \frac{m}{M}} \right) \cdot {g_k} + \frac{m}{M} \cdot {g_{k + 1}}
  \label{eq6}
\end{equation}
where ${g_k}\left( m \right)$ is the gain of the $m$-th frequency for the $k$-th band, ${g_k}$  and ${g_{k+1}}$ are the band gains for the $k$-th and the ($k+1$)-th bands, $M$ denoting the band length of the $k$-th band.

\subsubsection{Training of Double-Talk Detection}
\begin{figure}
\centering
\includegraphics[scale=0.48]{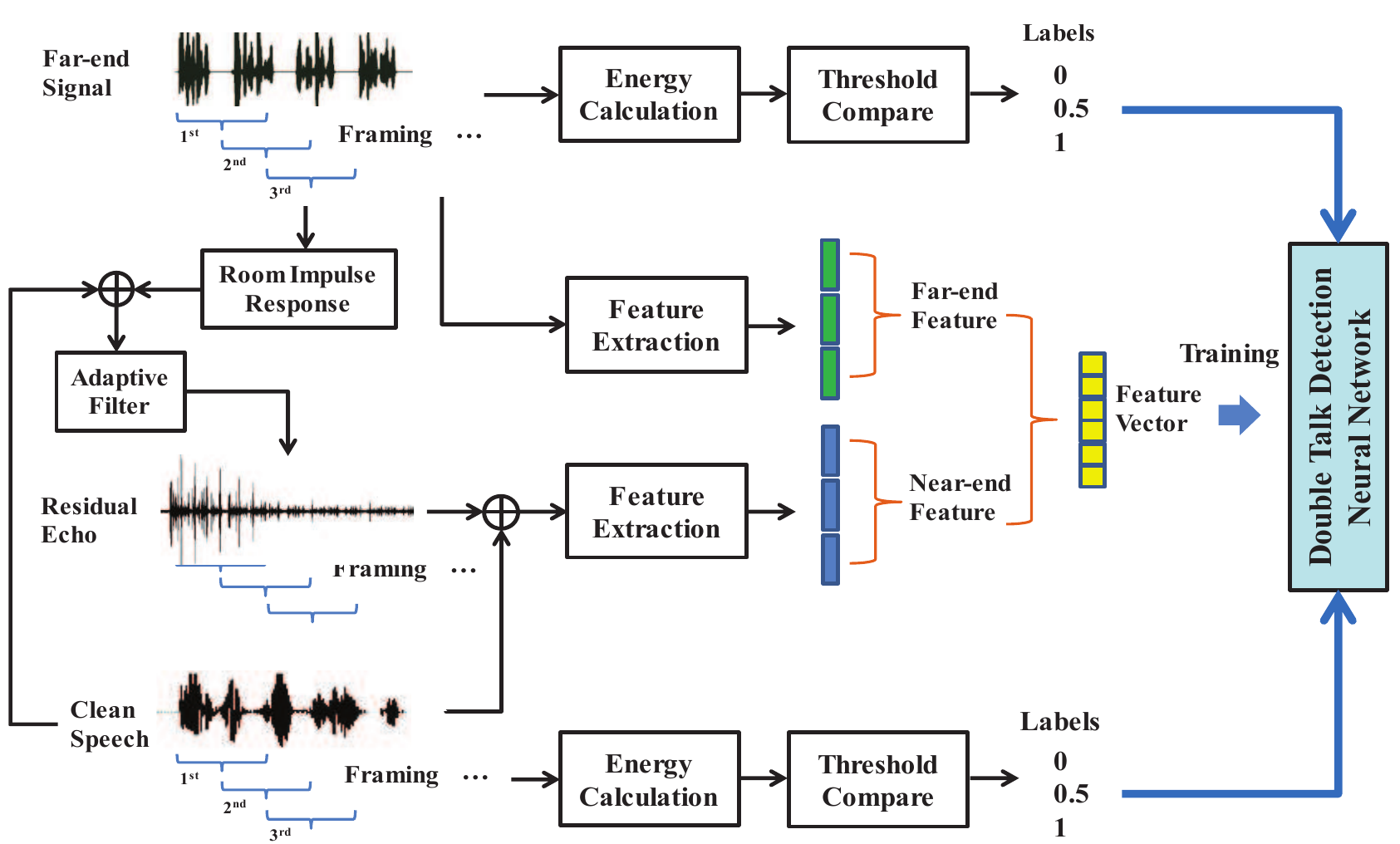}
\caption{Flowchart of double-talk detection training}
\label{fig:dtd}
\end{figure}

The training data can be either manually annotated or simulated. Manual labeling data is obtained by listening whether there is audio either at the far-end or the near-end and where there are. The audio file is recorded by an audio recording device where the audio played by the source superimposed by the one from the device itself is recorded. However, this method is time consuming. Therefore, simulation data is used for training. This can be illustrated in Fig. \ref{fig:dtd} and be summarized as follows:
\begin{enumerate}
\item \textbf{Far-end data preparation.} Far-end data is the reference signal for echo cancellation which is the audio file transmitted in the reference channel before playing out by the loudspeaker of the devices itself. This reference signal is framed and windowed, then used for energy calculation. This energy value is compared with two thresholds so that it is labelled by ``1'' if it is larger than the higher threshold, and labelled by ``0'' if it is lower than the lower threshold, otherwise, labelled by ``0.5''. This labels are calculated frame by frame, representing the probability of audio existing, together with feature vectors.

\item \textbf{Near-end data preparation.} Here, the near-end data is the signal after adaptive filtering where vast echo especially for the linear echo will be cancelled. As for the echo signal, it is obtained by convolving the reference signal with the RIRs. This echo signal is mixed with a clean speech audio file for simulating the microphone receiving signal. This microphone signal is then processed by adaptive filtering. Thereafter, clean speech mixed by residual echo is obtained, representing the near-end data for training. The labels representing whether there is clean speech or not can be easily obtained by directly calculating the energy and comparing with the thresholds. It is notable that since the amplitude of residual echo is relative small compared with that of clean speech, the labels can be also obtained by directly calculating the signal energy after adaptive filtering. Similarly, the feature vectors corresponding to each frame is calculated.

\item \textbf{Training process.} Since the labels for the two channels can be directly obtained by comparing the frame energy with thresholds, the voice detections can be implemented individually with VADs. With the feature vectors and their labels, the VAD modules for each channel can be trained without too much difficulty.
\end{enumerate}

\subsubsection{Training of Residual Echo Suppression}
\begin{figure}
\centering
\includegraphics[scale=0.5]{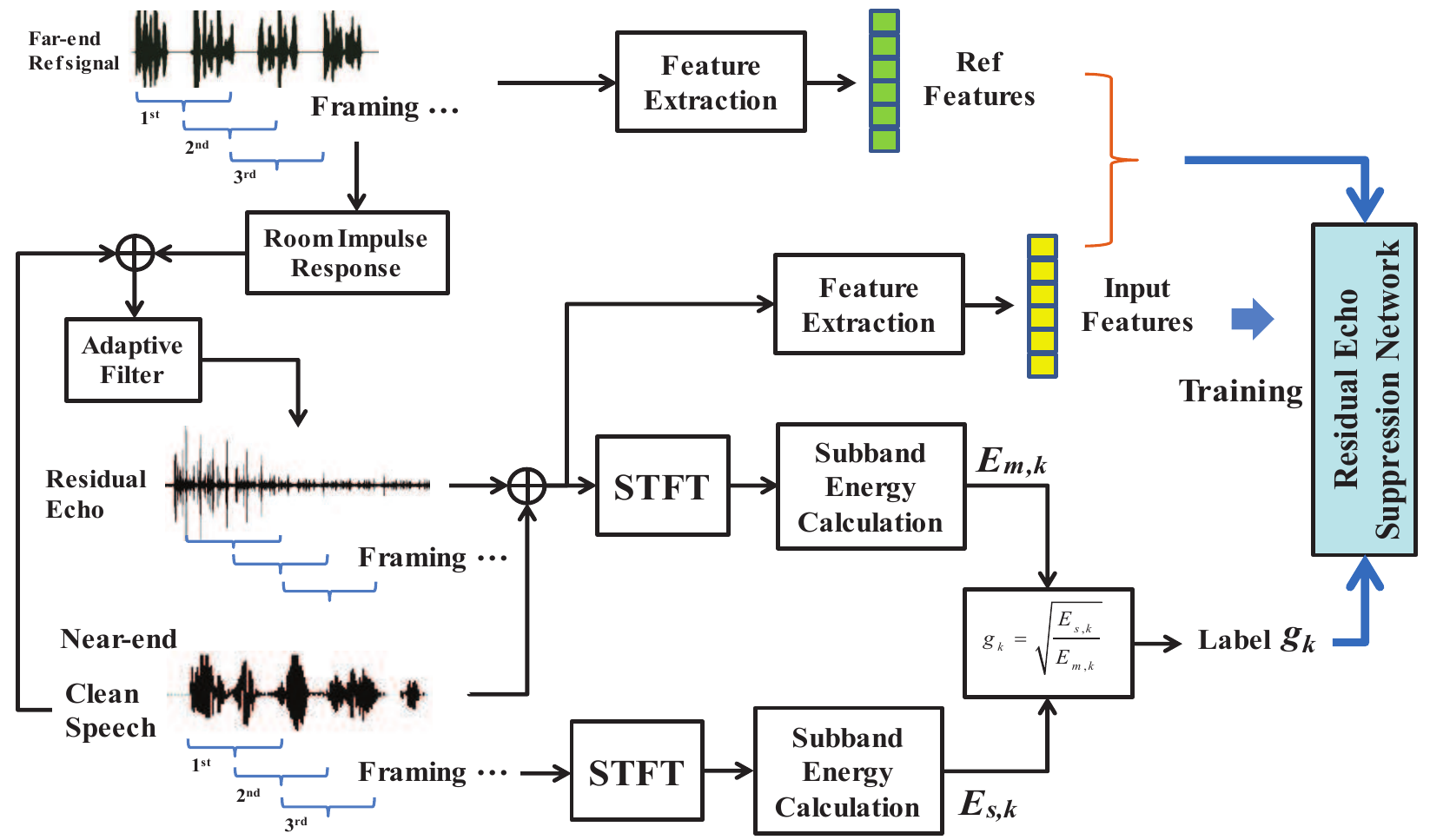}
\caption{Flowchart of frequency band gain training}
\label{fig:gain}
\end{figure}

The aim for residual echo suppression network is calculating the band gains whose training process is depicted in Fig. \ref{fig:gain}. The far-end and the near-end data are prepared in the same way as aforementioned except for the labels of band gains. These can be obtained by calculating the band energy of clean speech denoted by ${E_{s,k}}$ and that of the residual signal after adaptive filtering denoted by ${E_{m,k}}$, and then dividing them band by band for getting the labels, i.e., ${g_k} = \sqrt {\frac{{{E_{s,k}}}}{{{E_{m,k}}}}}$. Meanwhile, the feature vectors of these two channels are the same as aforementioned.

\section{Performance Evaluation}
\begin{figure}[H]
\centering
\includegraphics[scale=0.80]{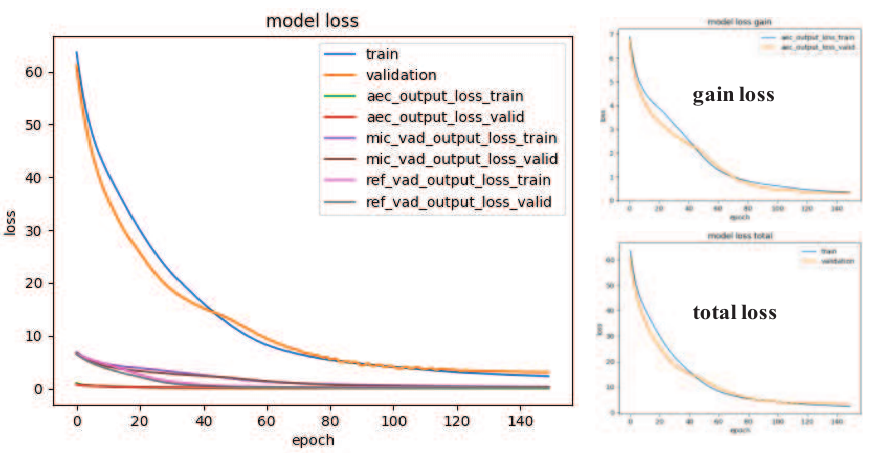}
\caption{Loss during training process}
\label{fig:loss_plot}
\end{figure}

$\bf{Model\ Training}$. The model structure can be shown in \ref{fig:structure}. A total of 10 hours of speech and 5 hours of echo data are constructed, resulting in 20 hours for training by using various combinations of gains and filters. In training process, three objective functions should be learned, i.e., the VAD of speech signal, the VAD of reference signal and the band gains for suppression. As is shown in Fig. \ref{fig:loss_plot}, both the training loss and the validation loss go down gradually approaching zero, revealing that a considerable model has been trained.

\begin{figure}[H]
\centering
\includegraphics[scale=0.70]{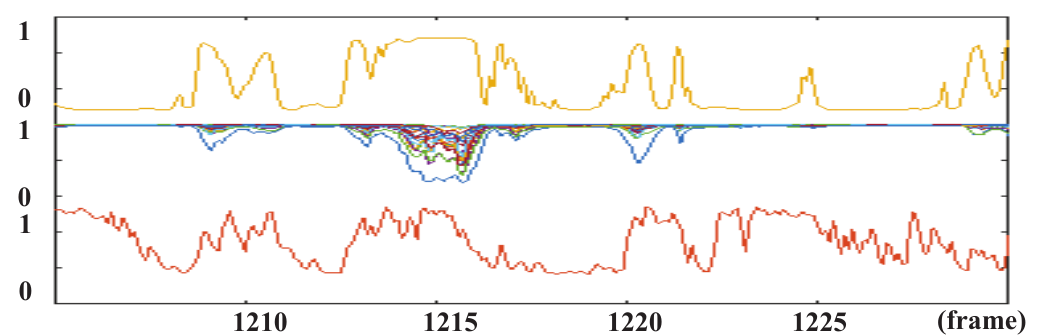}
\caption{Band gain calculated by neural network}
\label{fig:gain_plot}
\end{figure}

\begin{figure}[H]
\centering
\includegraphics[scale=0.65]{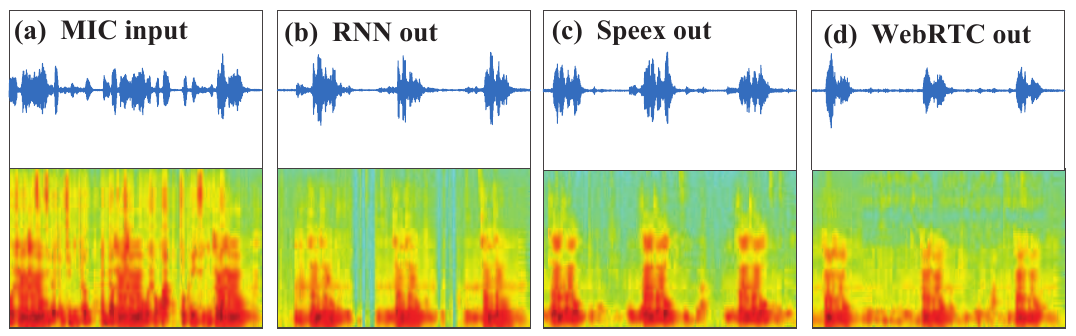}
\caption{Wave comparisons}
\label{fig:wavcomp}
\end{figure}

$\bf{Experiments\ Validation}$. a) \emph{Band Gain}. A piece of audio speech consisting by a string of wakeup words interfered by text to speech (TTS) audio played by itself are used for measurements. The calculated results about the VADs and the band gains are depicted in Fig. \ref{fig:gain_plot}. It can be found that, the band gains would approach zero if reference signal is detected at the momentum of wakeup word appearing. Since the energy of residual echo gathers at low bands, therefore the band gains for suppression would be lower for the low bands than that of the higher bands. b) \emph{Wave Observation}. For evaluating the performance, methods from prevailing open source codes are extracted for comparisons. These can be seen from Fig. \ref{fig:wavcomp} that the residual echo after the proposed RNN algorithm can be suppressed a lot compared with those of Speex and WebRTC. These are more obvious at the speech gaps where only residual echo exist. It can also be found that the spectrums at high bands are cut after WbeRTC AEC, which may be introduced by the non-linear processing (NLP) in the algorithm.

c) \emph{Performance Comparisons}. The ERLE value representing the echo suppression performance, and the LSD representing the spectrum loss in terms of the voice caused by AEC are evaluated and listed in Table~\ref{tab:performance}. Since AEC module is mostly implemented on devices, the response time (RT) obtained on the same platform representing the processing speed, and the model size representing the algorithm complexity should also be considered. It can be seen that the proposed scheme can obtain higher ERLE with considerable spectrum loss, processing time. Though the module size of the proposed scheme is larger, since the reference signal is clean speech, the model structure of VAD for this channel can be tailored. Meanwhile, the intermediate results of VADs for echo estimation in the model structure are likely to be clipped. These all could reduce the model size.

\begin{table}[H]
\centering
\caption{Comparisons by experiments}
  \label{tab:performance}
 \begin{tabular}{|c|c|c|c|c|}
   \hline
   Algorithms & ERLE & LSD & RT & Size \\
   \hline
    Speex & 25 dB & 1.01 dB  & 0.42 ms/frame & 106 kb\\
   \hline
    WebRTC & 40 dB & 1.66 dB  & 0.45 ms/frame & 140 kb\\
   \hline
   Proposed & 68 dB & 1.18 dB  & 1.63 ms/frame & 450 kb\\
   \hline
\end{tabular}
\end{table}

\section{Conclusions}
A combination scheme by concatenating adaptive filter and neural network is proposed for acoustic echo cancellation. The echo can be cancelled in a large scale after adaptive filtering, especially for linear echo, leaving the residual echo a bit. The spectrum of residue is much different compared with the speech audio, and can be considered as a special type of noise. Therefore, this residue is suppressed to a considerable level by employing neural network. Experiments reveal that the proposed scheme can obtain higher performance of echo suppression with considerable spectrum damage and response time.

\bibliographystyle{IEEEtran}

\bibliography{mybib}

\begin{thebibliography}{1}
\providecommand{\url}[1]{#1}
\csname url@samestyle\endcsname
\providecommand{\newblock}{\relax}
\providecommand{\bibinfo}[2]{#2}
\providecommand{\BIBentrySTDinterwordspacing}{\spaceskip=0pt\relax}
\providecommand{\BIBentryALTinterwordstretchfactor}{4}
\providecommand{\BIBentryALTinterwordspacing}{\spaceskip=\fontdimen2\font plus
\BIBentryALTinterwordstretchfactor\fontdimen3\font minus
  \fontdimen4\font\relax}
\providecommand{\BIBforeignlanguage}[2]{{%
\expandafter\ifx\csname l@#1\endcsname\relax
\typeout{** WARNING: IEEEtran.bst: No hyphenation pattern has been}%
\typeout{** loaded for the language `#1'. Using the pattern for}%
\typeout{** the default language instead.}%
\else
\language=\csname l@#1\endcsname
\fi
#2}}
\providecommand{\BIBdecl}{\relax}
\BIBdecl

\bibitem{echo}
E.~H$\ddot{a}$nsler, G.~Schmidt, ``Topics in acoustic echo and noise control: selected methods for the cancellation of acoustical echoes, the reduction of background noise, and speech processing,'' Springer Berlin Heidelberg, 2006.

\bibitem{aec}
J.~Benesty, T.~G$\ddot{a}$nsler, ``Advances in network and acoustic echo cancellation,'' \emph{Advances in network and acoustic echo cancellation}, Springer, 2001.

\bibitem{LMS}
E.~Ferrara, ``Fast implementations of LMS adaptive filters,'' \emph{IEEE Transactions on Acoustics, Speech, and Signal Processing}, vol.~28, no.~4, pp.~474--475 1980.

\bibitem{NLMS}
R.~Tyagi, R.~Singh and R.~Tiwari, ``The performance study of NLMS algorithm for acoustic echo cancellation,'' in \emph{International Conference on Information, Communication, Instrumentation and Control}, ICICIC, 2017, pp.~1--5, Indore.

\bibitem{BLMS}
G.~A.~Clark, S.~K.~Mitra, and S.~R.~Parker, ``Block implementation of adaptive digital filters,'' \emph{IEEE Trans. Acoust., Speech, Signal Processing}, vol.~ASSP--29, pp.~744--752, June 1981.

\bibitem{PBFDAF}
P¨¢ez Borrallo Jos¨¦M., M.~G.~Otero, ``On the implementation of a partitioned block frequency domain adaptive filter (PBFDAF) for long acoustic echo cancellation,'' \emph{Signal Processing}, vol.~27, no.~3, pp.~301--315, 1992.

\bibitem{MDF}
J.~S.~Soo, K.~K.~Pang, ``Multidelay block frequency domain adaptive filter,'' \emph{IEEE Transactions on Acoustics, Speech and Signal Processing}, vol.~38, no.~2, pp.~373--376, 1990.

\bibitem{rate}
J.~Valin, ``On adjusting the learning rate in frequency domain echo cancellation with double-talk,`` \emph{IEEE Transactions on Audio, Speech, and Language Processing}, vol.~15, no.~3, pp.~1030--1034, 2007.

\bibitem{vss1}
Z.~Yuan and X.~Songtao, ``Application of new LMS adaptive filtering algorithm with variable step size in adaptive echo cancellation,'' in \emph{IEEE International Conference on Communication Technology}, ICCT, 2017, pp.~1715--1719.

\bibitem{vss2}
J.~Benesty, H.~Rey, L.~R.~Vega and S.~Tressens, ``A nonparametric VSS NLMS algorithm,'' \emph{IEEE Signal Processing Letters}, vol.~13, no.~10, pp.~581--584, 2006.

\bibitem{vss3}
C.~Paleologu, S.~Ciochina and J.~Benesty, ``Double-talk robust VSS-NLMS algorithm for under-modeling acoustic echo cancellation,'' in \emph{IEEE International Conference on Acoustics, Speech and Signal Processing}, ICASSP, 2008, pp.~245--248.

\bibitem{vss4}
Mohammad Asif Iqbal and S.~L.~Grant, ``Novel variable step size nlms algorithms for echo cancellation,'' in \emph{IEEE International Conference on Acoustics, Speech and Signal Processing}, ICASSP, 2008, pp.~241--244.

%

\bibitem{NLP3}
O.~Tanrikulu and K.~Dogancay, ``A new non-linear processor (NLP) for background continuity in echo control,'' in \emph{IEEE International Conference on Acoustics, Speech, and Signal Processing}, ICASSP, 2003, pp.~V--588.

\bibitem{NLP4}
M.~Doroslovacki, ``Optimal non-linear processor control for residual-echo suppression,'' in \emph{IEEE International Conference on Acoustics, Speech, and Signal Processing}, ICASSP, 2003, pp.~V--608.

\bibitem{NLP5}
B.~Panda, A.~Kar and M.~Chandra, ``Non-linear adaptive echo supression algorithms: A technical survey,'' in \emph{International Conference on Communication and Signal Processing}, ICCSP, 2014, pp.~076--080.

\bibitem{kalman}
M.~Z.~Ikram, ``Non-linear acoustic echo cancellation using cascaded Kalman filtering,'' in \emph{IEEE International Conference on Acoustics, Speech and Signal Processing}, ICASSP, 2014, pp.~1320--1324.

\bibitem{modeling}
M.~I.~Mossi, C.~Yemdji, N.~Evans, and etc., ``Robust and low-cost cascaded non-linear acoustic echo cancellation,'' in \emph{IEEE International Conference on Acoustics, Speech and Signal Processing}, ICASSP, 2011, pp.~89--92.

\bibitem{RIR}
J.~Mourjopoulos, ``On the variation and invertibility of room impulse response functions,'' \emph{Journal of Sound $\&$ Vibration}, vol.~102, no.~2, pp.~217--228, 1985.

\bibitem{speex1}
J.~M.~Valin, ``Speex: A free codec for free speech,'' \emph{Speex A Free Codec for Free Speech}, 2016.

\bibitem{speex2}
P.~Srivastava, K.~Babu and T.~Osv, ``Performance evaluation of Speex audio codec for wireless communication networks,'' in \emph{International Conference on Wireless and Optical Communications Networks}, WOCN, 2011, pp.~1--5.

\bibitem{RNN}
Valin, Jean-Marc, ``A hybrid DSP/deep learning approach to real-time full-band speech enhancement,'' in \emph{IEEE International Workshop on Multimedia Signal Processing}, MMSP, 2018, pp.~1--5.


\end{thebibliography}


\end{document}